\documentclass[conference]{IEEEtran}
\IEEEoverridecommandlockouts
\usepackage{cite}
\usepackage{amsmath,amssymb,amsfonts}
\usepackage{algorithmic}
\usepackage{graphicx}
\usepackage{textcomp}
\usepackage{xcolor}
\def\BibTeX{{\rm B\kern-.05em{\sc i\kern-.025em b}\kern-.08em
    T\kern-.1667em\lower.7ex\hbox{E}\kern-.125emX}}
\begin{document}

\title{Efficient Laser Frequency Allocation in Packet-Optical Nodes with Coherent Transceivers\\
}

\author{\IEEEauthorblockN{Constantine A. Kyriakopoulos}
\IEEEauthorblockA{
Barcelona, Spain \\
orcid.org/0000-0001-7874-2205\\ 
kyriak@csd.auth.gr\\
}
}

\maketitle

\begin{abstract}
The introduction of silicon chipsets with the capability of processing incoming optical packet traffic, creates a new generation of packet-optical nodes, the whiteboxes. Their inherent functionality of carrying pluggable coherent transceiver modules, extends their scope in the field of transport optical networks. Also, their hybrid nature improves the overall efficiency since higher layer functionality is now performed at wire speed. This is feasible by embedding in the computational logic, apart from the typical packet inspection routines and security features, Machine Learning-based traffic engineering as well. In this work, a topology based on multiple packet-optical nodes residing at the edges (ingress and egress) of a ROADM network, is evaluated according to the average laser frequency configuration time. This is achieved by exploiting telemetry analytics which are collected by the CMIS driver of the pluggable transceivers, for allocating efficient frequency slots to the source and destination of connectivity requests traversing through the ROADM network. This pair of nodes is supervised by the packet SDN controller which is part of the control plane of the optical transport network. This controller receives telemetry feedback from the whiteboxes and uses it to execute efficient ML techniques locally, for finding efficient frequency slots for the incoming transport requests. Next, it applies them to request's edge nodes. The decrease of the average laser configuration time is achieved in the evaluated topology, improving the overall efficiency of the control plane.
\end{abstract}


\section{Introduction}

Optical switching undergoes significant performance improvement since the dawn of dedicated silicon Application-Specific Integrated Circuits (ASICs) \cite{sun2020800g} during the last few years. Switching is now a fast process executing at the wire speed-level. Previously, the main system processor along with the full networking stack was executing the switching routines. This caused significant decrease to the overall performance, considering also the case of taking multiple routing decisions simultaneously.

In this new era, a domain-specific programming language controls the packet forwarding planes in networking devices. This is the Programming Protocol-independent Packet Processors (P4) \cite{hauser2023survey} language. Programmes are designed to be implementation-independent, i.e., they can be compiled for different hardware vendors and execute literally from within the ASICs. It includes the constructs that are optimised for network data forwarding. Even small neural networks are demonstrated to perform well at the wire speed designed specifically with P4 in their logic \cite{cugini2023telemetry, goswami2023survey}.

Whiteboxes \cite{scano2021hierarchical, lopez2020enabling} are getting a lot of attention lately. These are packet-optical nodes suitable for the edges of transport networks. They offer abstraction layers for designing the switching processes in an open and vendor-independent way. These are advanced switches including the aforementioned technologies over an underlying operating system, like SONiC OS \cite{scano2021hierarchical} or OcNOS \cite{berde2014onos}. For example, configuration and access to the pluggable devices \cite{sgambelluri2021coordinating} and the switching capabilities is performed with a Common Management Interface Specification (CMIS) driver interface or the vendor-agnostic Switch Abstraction Interface (SAI) API. Below this API, resides a vendor-specific ASIC SDK which uses vendor-locked kernel drivers as modules. Using this design, OS releases can come with different hardware configurations, having their main processes implementation-agnostic, facilitating this way the interoperability between different platforms.

Coherent module drivers are also standardised by using the CMIS specification. In this case, management-related communication is based on an interface, such as QSFP Double Density (QSFP-DD), OSFP, and others. It is an open standard for system manufacturers, integrators, and distributors of transceivers and similar equipment. Network operating systems targeting switching devices come with full CMIS support over the last few years.

In this work, considering the aforementioned developments, the coherent pluggable average laser configuration time is improved while the process is evaluated according to a testbed setup. This is facilitated in the appropriate lab equipment (or via simulation when required), on a typical control plane of a ROADM metro network scenario which consists of multiple whiteboxes at its edges. For that purpose, a software agent (Netconf server) residing at the whiteboxes, is developed receiving input from the Software-Defined Networking (SDN) packet controller (PacketCTL - a Netconf client). Then, configuration of the local transceiver laser frequencies of the controlled pluggable devices takes place, for facilitating the connectivity in-between the ROADM network. Also, the agent records and reports back telemetry data (feedback) which is used by the PacketCTL's resource-allocating mechanism to improve efficiency within the network topology.

PacketCTL is able to improve the overall control plane configuration time, by utilising an internal Feed-forward Neural Network (FNN) which is based on Q-Learning (QL). This Reinforcement Learning (RL) mechanism adapts to the current operating environment by efficiently allocating frequency slots to the pluggables, decreasing the average laser configuration times. To achieve this, it exploits the previous configuration times as feedback input while in learning state.

The decrease of the average configuration time of transceiver laser frequencies is achieved and lies in the range of 20 - 25\%. This contributing factor in the control plane of the current research is important for the future transport optical networks, since whiteboxes will have an important role in 6G connectivity. Also, this adaptive logic can extend to the ROADM part of the network as well. Different paths through the ROADM devices are configured more efficiently when exploiting the operating conditions in a similar RL fashion that executes at the Optical Controller (OptCTL) premises.

The next sections are organised as follows. Related research is next (Section \ref{lab:related}). Agent functionality is elaborated upon in Section \ref{lab:agent}. Next, the network environment is described in detail (Section \ref{lab:topo}), and the results that show the efficiency of the RL logic follow in the last Section \ref{lab:res}.

\section{Related Research\label{lab:related}}

MANTRA (Metaverse ready Architectures for Open Transport) is a group \cite{de2022mantra} with the goal of constructing an end-to-end reference network architecture. This facilitates the transition from aggregated to vendor-independent disaggregated open network architectures. As an implicit result, a new generation of IPoWDM networks with switches carrying 400G coherent pluggable transceivers is emerging rapidly. This architecture lies in the core of the current research which enables the efficient communication between SDN controllers and core optical routing devices (Figure \ref{figEdgecore}).

Following the open initiatives, vendor lock-in concerning transceiver devices is eliminated \cite{sgambelluri2020openconfig} by utilising standards like OpenConfig and OpenROADM. Transmission modes, and others attributes as well, are exposed with YANG models \cite{szyrkowiec2017optical}, so the SDN controller can deploy algorithms that aim at efficient mode selection during network's operational state. This enables the design of software agents handling the communication between entities in the control plane, similar to the current research.

Relying on MANTRA, conventional functionality can adapt to the future network requirements. Failure recovery is important aspect for adoption since it is not strictly defined in this standard and increases the quality of service. Some related methods aiming at failure recovery and can be used within MANTRA are presented and evaluated \cite{sgambelluri2024failure}. Tuning performance of pluggable coherent devices is evaluated as well, along with recovery time with real traffic and estimation of the channel bandwidth.

Effective cooperation of packet-optical nodes which are managed by the BGP and OSPF protocols and a hierarchical control architecture, is demonstrated \cite{scano2023hybrid}. It aims at the orchestrated provisioning and soft failure recovery in a metro network topology. The presented perspective is based on a variation from MANTRA where the OptCTL configures the optical domain and the coherent pluggables, while the PacketCTL configures the BGP and OSPF instances of the packet-optical nodes.

SDN solutions with communicating entities (like in current research) which coordinate and control modern pluggable transceivers in a multi-layer network consisting of whiteboxes are presented \cite{giorgetti2023enabling}. These two approaches, i.e., the exclusive and the shared one, are evaluated according to defined workflow experiments. Included in the results are, the end-to-end connection setup time, the optical intent setup time, the discovery time of the new link in the packet domain, and finally, the packet intent setup time.

\section{Agent and Controller Functionality\label{lab:agent}}

PacketCTL (Netconf client) receives commands from the Hierarchical Controller (HrCTL) consisting of a pair of nodes, i.e., the ingress and egress whitebox of the transport route. Its purpose is to choose efficient frequency slots on both the edges, so as to minimise the average configuration time of the control plane. To achieve this goal, an internal ML process executes which is based on RL, using a QL method to evaluate the available frequency slots in each request. To adapt while staying in the operating network state, it utilises the feedback that is returned from the whiteboxes via the Netconf protocol. This is the negative laser configuration time of each successfully set frequency slot on the pluggable device. This way, values leading to zero evaluate the slots higher in ranking, so there are more chances for those to be chosen during subsequent requests.

\begin{figure}[btp]
\centerline{\includegraphics[width=0.5\textwidth]{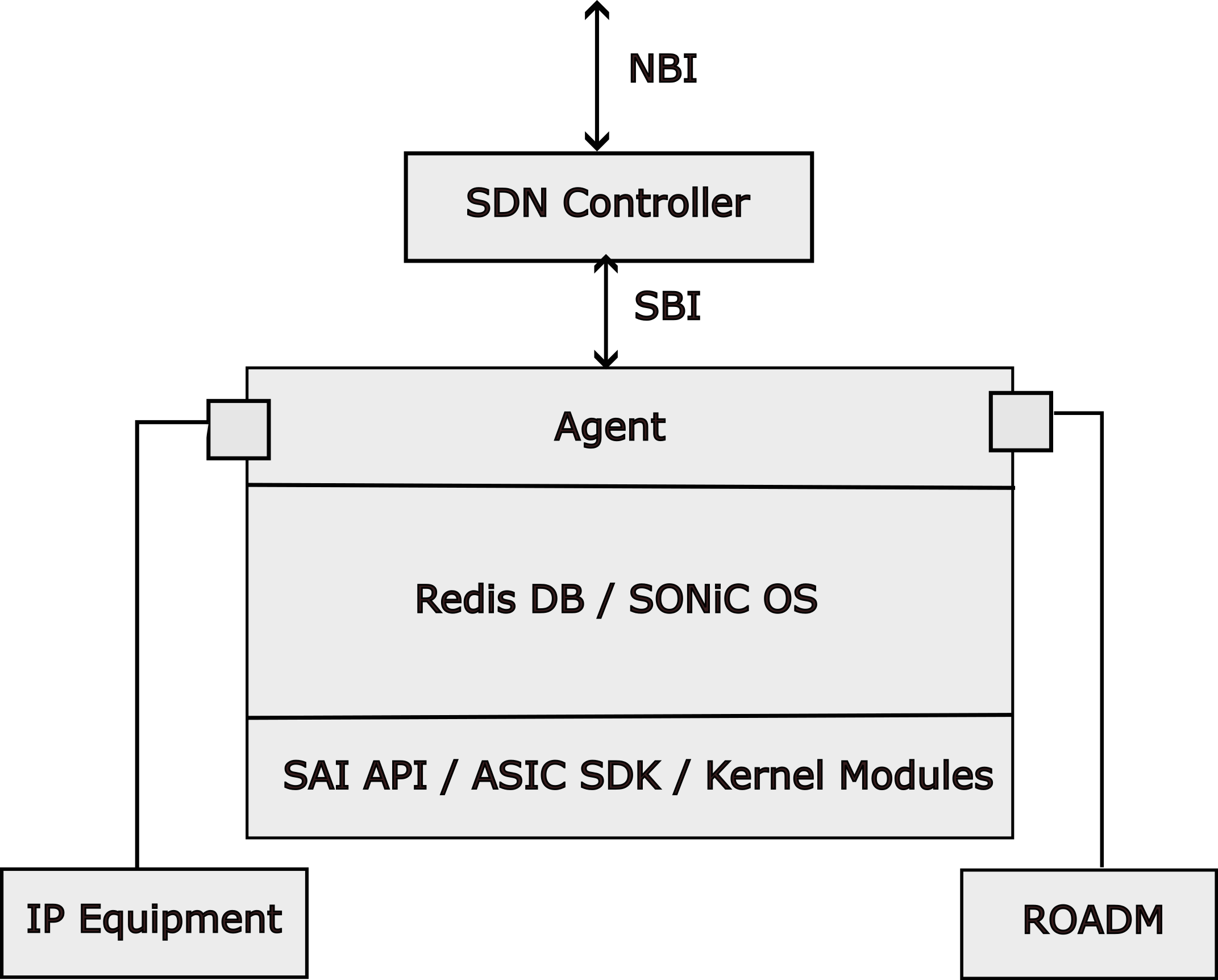}}
\caption{Packet/optical node.}
\label{figNode}
\end{figure}

A software agent (Netconf server) sits within a docker container in the application space of SONiC OS, having the capability to communicate with the South Bound Interface (SBI) of the PacketCTL, and at the same time, to configure the pluggable parameters via the RedisDB which the SONiC architecture relies upon (Figure \ref{figNode}). Also, in case a P4 \cite{p4standard} chipset is present, it can execute resource allocation routines at hardware speed. The ASIC chipset is also capable of performing switching functionality for the purpose of bypassing the main CPU of the system to boost performance. The agent functions differently by fetching input from the SBI of the PacketCTL, using the Netconf protocol \cite{dallaglio2017control} with a YANG model. Parsing the input, the agent is able to alter the local RedisDB entries accordingly, so the SONiC OS daemons are able to detect the changes and execute routines for configuring the state of the pluggables, reporting telemetry data back via the system logs.

Figure \ref{figNode} depicts the packet/optical node. The SDN controller uses both the North Bound Interface (NBI) and SBI. The agent communicates with the other standard SONiC processes through the core of the system, i.e., the RedisDB. System daemons (residing at different docker containers) detect the altered values of the database and perform their functionality accordingly. For example, if the laser frequency value of a pluggable device in a specific port is altered, the daemon uses the SWSS API. This API communicates with the database container which is monitored by the syncd container for changes. The latter uses the proprietary kernel drivers to achieve its low level functionality on pluggables. The agent is connected to other IP equipment with Ethernet interfaces, and as well as to the ROADM part of the network as depicted in this figure.

\begin{figure}[btp]
\centerline{\includegraphics[width=0.5\textwidth]{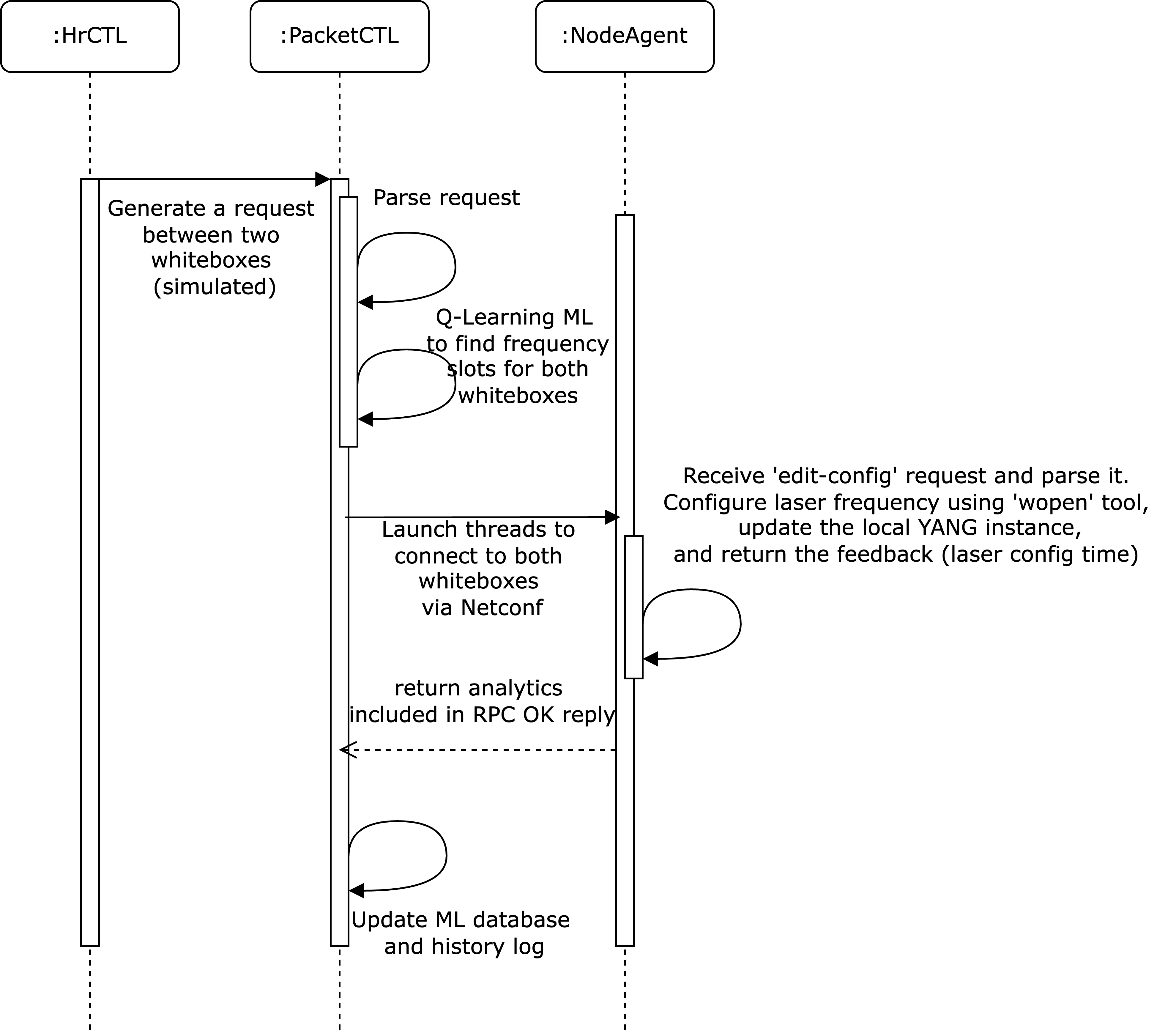}}
\caption{Sequence diagram.}
\label{figSequence}
\end{figure}

The procedure of setting a specific laser frequency on a pluggable device is depicted as a sequence in Figure \ref{figSequence}. Initially, a request is generated from the HrCTL, including a pair of whiteboxes. The local DB of the PacketCTL contains previous feedback values (configuration times) per transceiver ID in each whitebox. A RL method based on QL is applied and the transceiver frequency slots are chosen to fullfil the request. Next, it launches system threads to connect concurrently to both whiteboxes via the Netconf protocol, using an XML RPC \textit{edit-config} structure that contains the pluggable ID and the chosen slot. After successful configuration, the whitebox returns embedded in the RPC XML OK message, the actual configuration time as feedback. Finally, the controller updates its local DB accordingly.

Service and device data models with definitions and reference implementations are preferably open and vendor-independent. To achieve this goal, these must be specified in a modelling language, e.g., YANG. This leads to the choice of Netconf which is a widely used protocol complying to these requirements. Design and agent implementation adheres to these models to provide interoperability and continuous standardised development.

\section{Network Topology and Testbed Setup\label{lab:topo}}

\begin{figure}[btp]
\centerline{\includegraphics[width=0.49\textwidth]{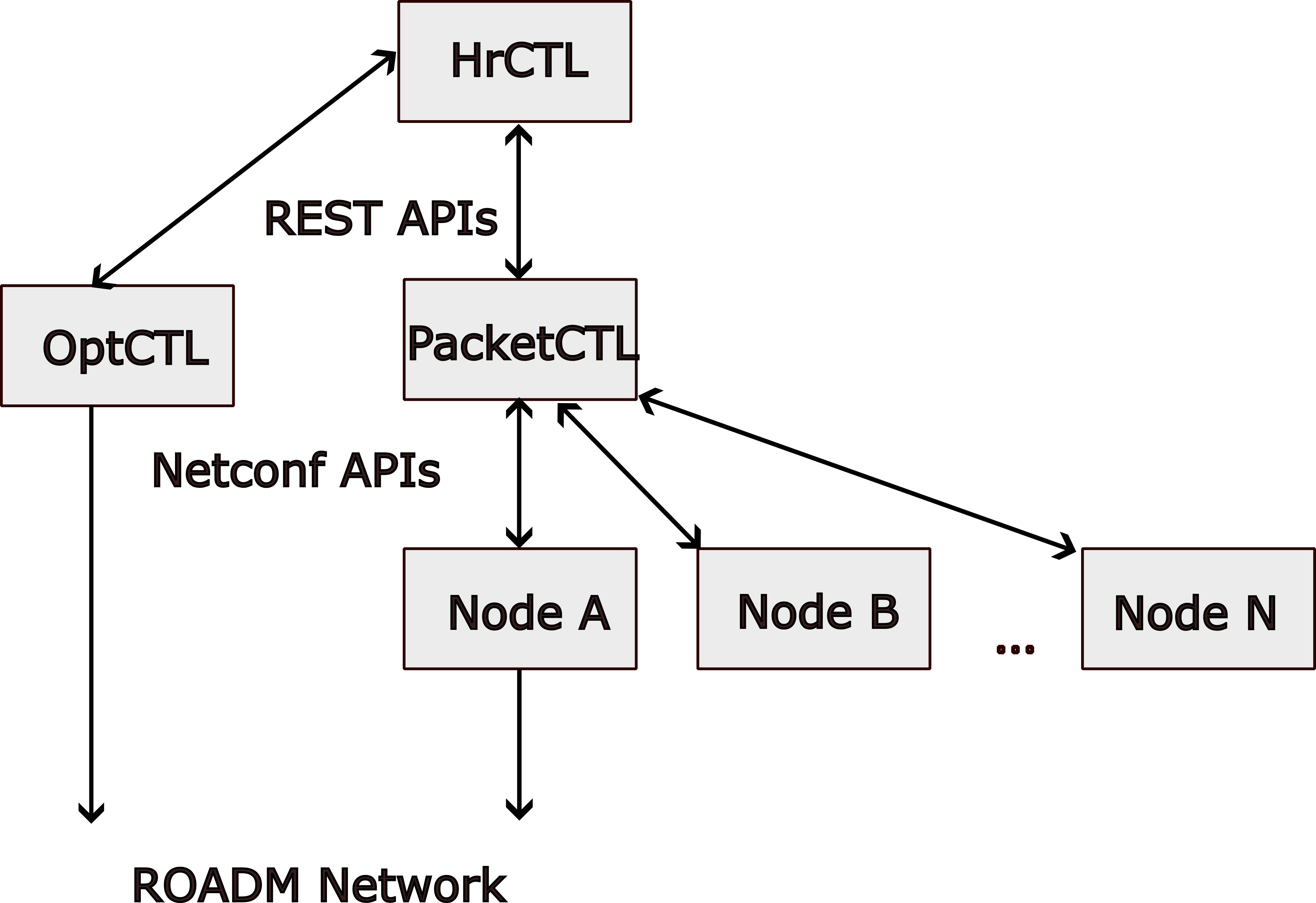}}
\caption{System connectivity.}
\label{figConnectivity}
\end{figure}

PacketCTL connects to multiple whitebox nodes concurrently (Figure \ref{figConnectivity}). These are the ingress and egress nodes of a ROADM network in-between. Commands arrive from a simulated HrCTL to the PacketCTL consisting of a pair of these nodes. PacketCTL runs locally a trained neural network to find frequency slots to allocate to each pluggable of both nodes. Each slot requires a slightly different laser configuration time to be enabled. So, the NN decreases the average configuration time on the whiteboxes by finding the most efficient frequency slots. It connects to the whiteboxes concurrently launching multiple threads (with Netconf and a designed YANG scheme for this purpose), using system sockets. The OptCTL configures the ROADM part of the network by enabling the path that reduces the resource usage and the propagation time, though this controller functionality is subject to a future research extension.

\begin{figure}[bp]
\centerline{\includegraphics[width=0.5\textwidth]{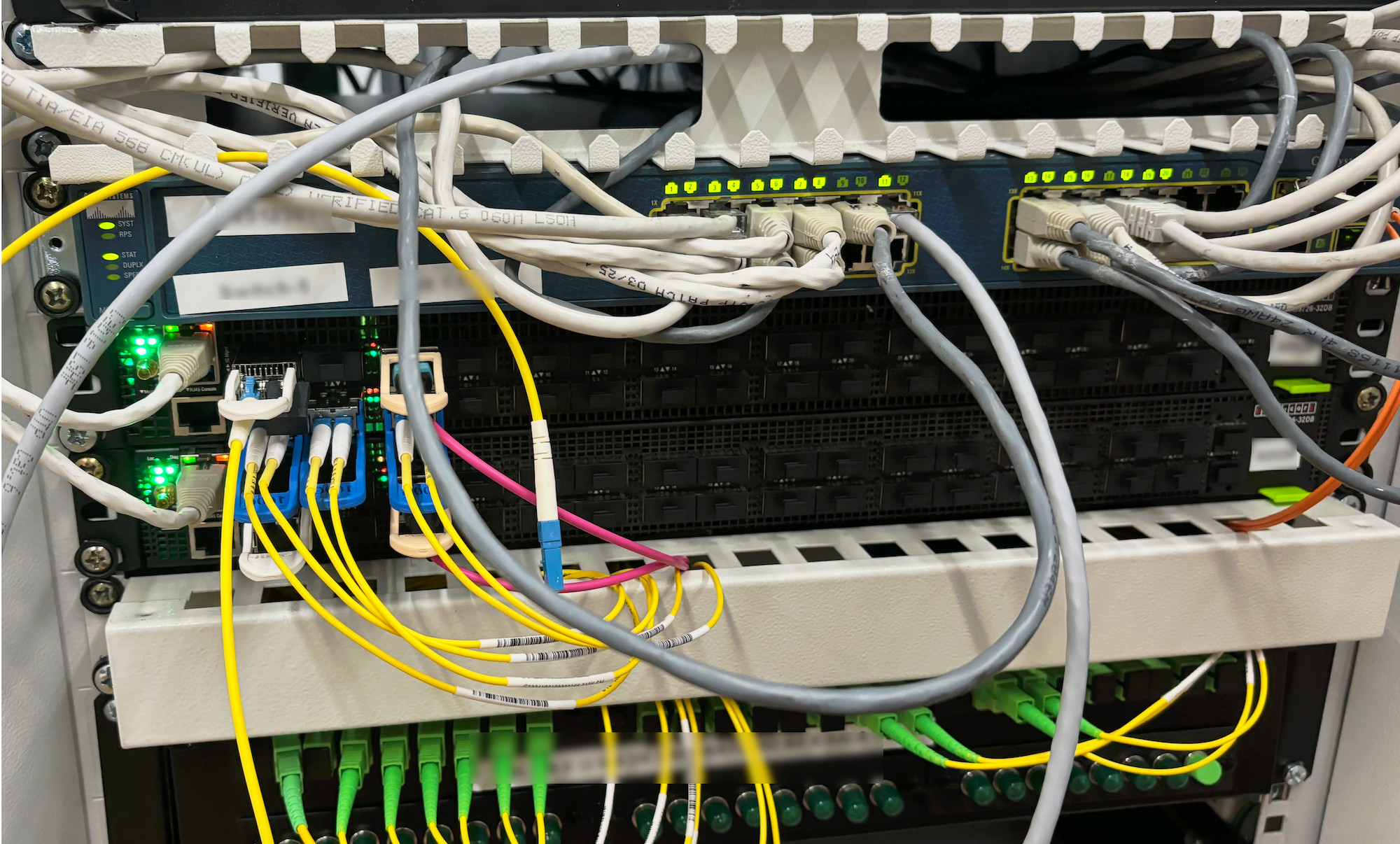}}
\caption{Two Edgecore DCS240 switches with Finisar ZR400-OFEC-16QAM pluggables.}
\label{figEdgecore}
\end{figure}

\begin{figure}[btp]
\centerline{\includegraphics[width=0.5\textwidth]{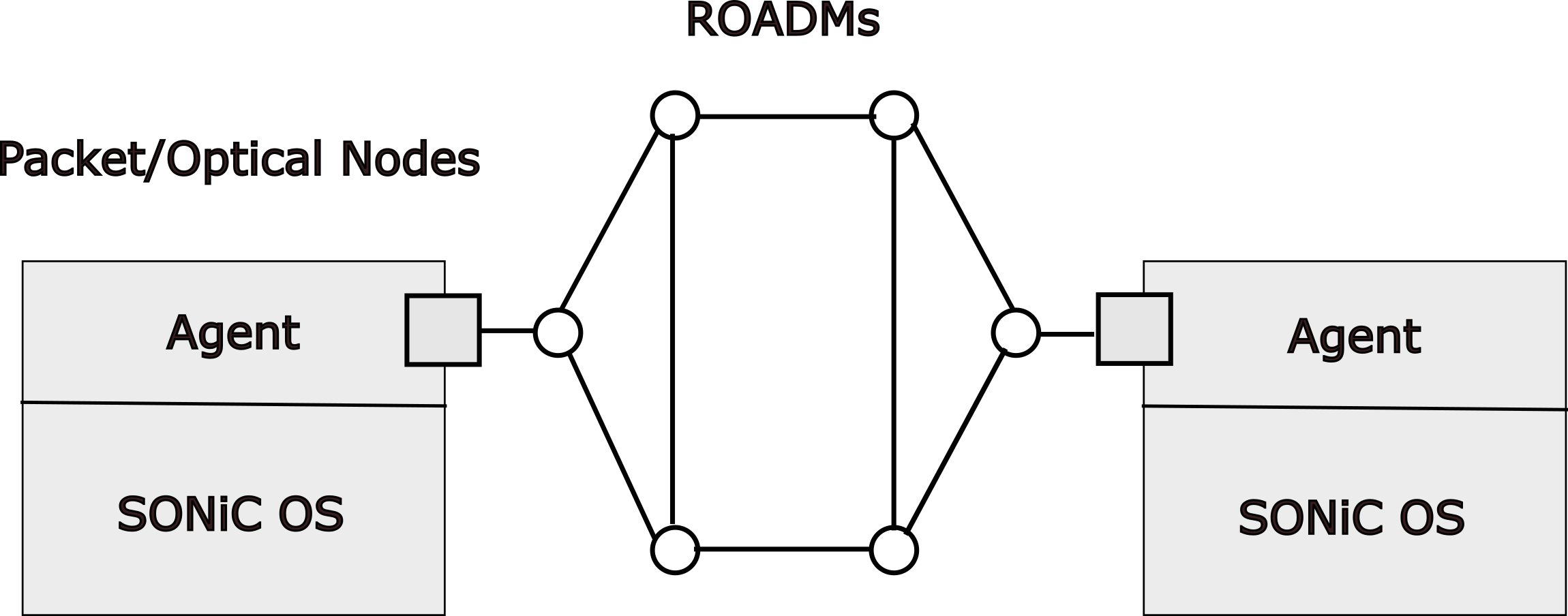}}
\caption{Request recipients.}
\label{figTopo}
\end{figure}

The lab equipment facilitating this project, consists of two Edgecore DCS240 AS9726-32DB switches (Figure \ref{figEdgecore}) relying on a Broadcom Trident 4 chipset to perform switching functionality. Also, ROADMs shape the metropolitan network (Figure \ref{figTopo}) that resides in between the switches (not part of the current research). Both whiteboxes run a specialised version of SONiC OS 202111.7 that comes equipped with the latest versions of the SAI API, the Broadcom SDK (supporting Trident 4) and CMIS drivers for usage with the Finisar ZR400-OFEC-16QAM pluggables (Figure \ref{figEdgecore}).

In Figure \ref{figTopo}, a symmetrical shape of the topology is depicted. The agent sits on both nodes at the application layer of SONiC OS (inside dockerised containers). ROADMs intermediate between the switch connectivity and can be configured by the OptCTL. Nodes are controlled as well by the corresponding PacketCTL. The propagation delay is dependent on the ROADM part of the network. Performing adaptive resource allocation in this part, mostly by utilising telemetry data at the node premises, leads to performance improvement.

Both switches operate on the following SONiC OS system information which includes the latest Broadcom proprietary ASIC software. Laser configuration times may depend upon specific software SDK versions since rapid development takes place in this field and system software updating is required to exploit the full hardware capabilities and potential.

{\footnotesize
\begin{verbatim}
SONiC Software Version: SONiC.Edgecore-SONiC
_20231012_082904_ec202111_hsdk_6.5.23_499
Distribution: Debian 11.8
Kernel: 5.10.0-8-2-amd64
Build commit: 2a98cb721
Build date: Thu Oct 12 18:35:06 UTC 2023
Built by: ubuntu@ip-10-5-1-65
Platform: x86_64-accton_as9726_32d-r0
HwSKU: Accton-AS9726-32D
ASIC: broadcom
ASIC Count: 1
\end{verbatim}}

Next, the logging output of the CMIS driver for setting the laser frequency at 192500 GHz and grid space at 100 GHz is displayed. The difference between the starting state of 'Datapath reinit' log entry until the timestamp of 'Ethernet0 configured...', is the laser configuration feedback time that the QL FNN exploits to adapt by evaluating its available choices. Each frequency slot uses a range of delay time values according to the hardware capabilities.

{\footnotesize
\begin{verbatim}
Jun 20 12:30:41.069151 sonic NOTICE pmon#xcvrd: 
CMIS: Ethernet0: force Datapath reinit
Jun 20 12:30:41.084331 sonic NOTICE pmon#xcvrd: 
CMIS: Ethernet0: 400G, lanemask=0xff, state=DP_DEINIT, 
appl=1, retries=0
Jun 20 12:30:44.478394 sonic NOTICE pmon#xcvrd: 
CMIS: Ethernet0: 400G, lanemask=0xff, state=DP_DEINIT, 
appl=1, retries=0
Jun 20 12:30:44.478394 sonic NOTICE pmon#xcvrd: 
CMIS: Ethernet0: 400G, lanemask=0xff, state=
AP_CONFIGURED, appl=1, retries=0
Jun 20 12:30:44.516702 sonic WARNING pmon#xcvrd: 
CMIS: Ethernet0 Tuning in progress, channel selection 
may fail!
Jun 20 12:30:44.582824 sonic NOTICE pmon#xcvrd: 
CMIS: Ethernet0 configured laser frequency 192500 
GHz grid space 100 GHz

\end{verbatim}}

\section{Results\label{lab:res}}

Both the Netconf entities (server and client) are designed and implemented in Python. The YANG structure is supported via the \textit{pyangbind} library which generates Python bindings (classes) from YANG models. The process that sets the actual laser frequencies is written in C++ and it is used to create the dataset for training the FNN. The functionality of this tool is also simulated for the cases where the equipment is not adequate to capture results such as when scaling the topology to 16 transceivers. All results are taken using a MacOS system with ARM M2 processor and 8GiB of memory. All processes are written with Python 3.12.x and Apple Clang++ 16.x.

For the scenaria of the network while it is operating, the module enabling new frequency slots on pluggables, produces output by using the average laser configuration time per frequency slot (real values from the dataset) and its standard deviation using a log-normal distribution. This way, it simulates the SONiC OS method of setting laser frequencies and the aforementioned C++ tool.

The dataset that trained the NN consists of real values that were recorded from one of the whiteboxes according to its CMIS system log. All available 49 slots (from 191300 to 196100 GHz) were configured for that purpose twice. The dataset is extended artificially with more values, so it includes the original ones plus synthetic augmented measurements. The latter consist of additional data points generated by adding Gaussian noise to the original measurements. Also, it retains the same slot numbers and introduces variability in configuration times for robustness. To record the real pluggable configuration times, a specially designed software tool is utilised (the C++ tool) which configures laser frequencies by altering the SONiC OS RedisDB values. Then, the related \textit{xcvrd} process (daemon) implements the changes upon the actual hardware and reports back to the system logs.

\begin{figure}[btp]
\centerline{\includegraphics[width=0.49\textwidth]{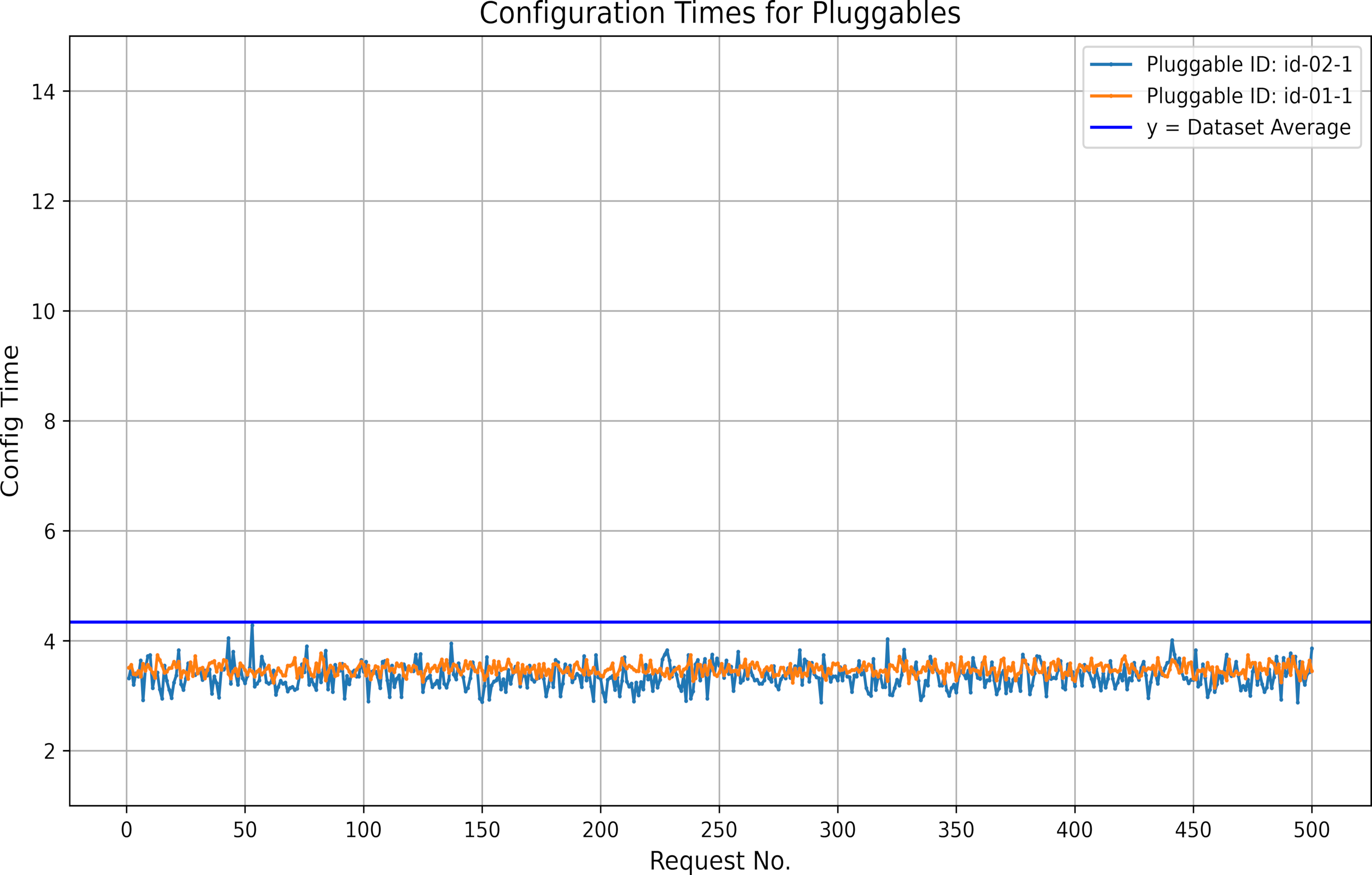}}
\caption{Configuration time of pluggables.}
\label{figFeedback}
\end{figure}

In Figure \ref{figFeedback}, the average laser configuration time from the dataset is represented by the horizontal blue line at y = 4.34 sec. For that purpose, 500 requests coming from the HrCTL are generated (x-axis). Both pluggable devices have average configuration times between 3.2 and 3.6 sec which is an improvement in the range of 20-25\% in comparison to the dataset’s average. This is the operating state (after the initialisation phase with training) of the topology which decreases the average configuration time of the control plane as requests are fulfilled one by one.

\begin{figure}[btp]
\centerline{\includegraphics[width=0.49\textwidth]{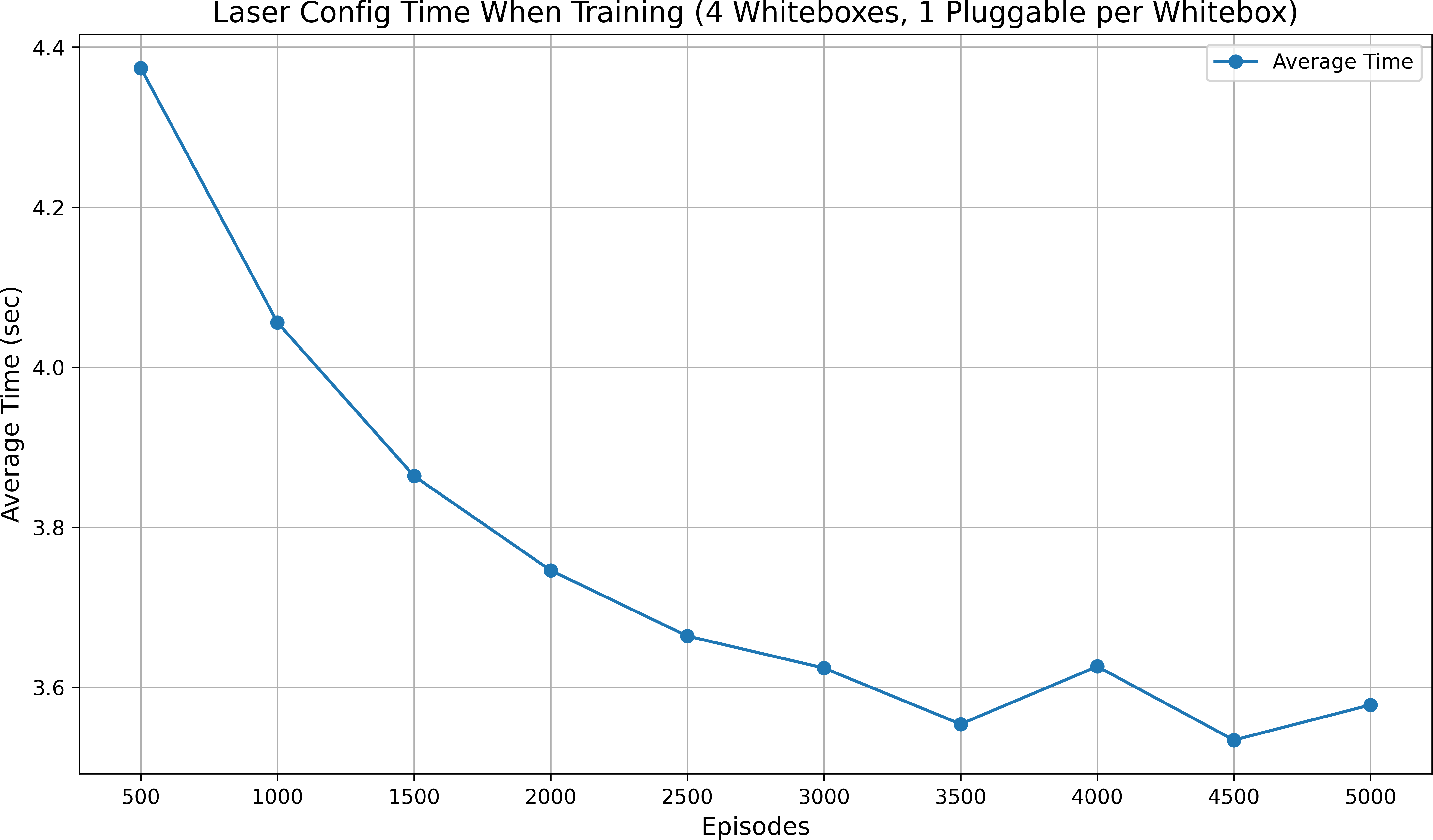}}
\caption{Training the ML agent.}
\label{figTraining}
\end{figure}

In Figure \ref{figTraining}, the training of the ML mechanism is depicted. It starts with high average configuration time, and as the number of episodes increases, it tends to reach an efficient value. This training process is recorded from 4 whiteboxes running in parallel (1 pluggable device each). After 3K episodes, the ML module reaches its steady state. Computing resources are minor since the extent of the dataset (with the supported frequency slots) is limited. This training decreases the average configuration time later on, while in operation mode.

\begin{figure}[btp]
\centerline{\includegraphics[width=0.49\textwidth]{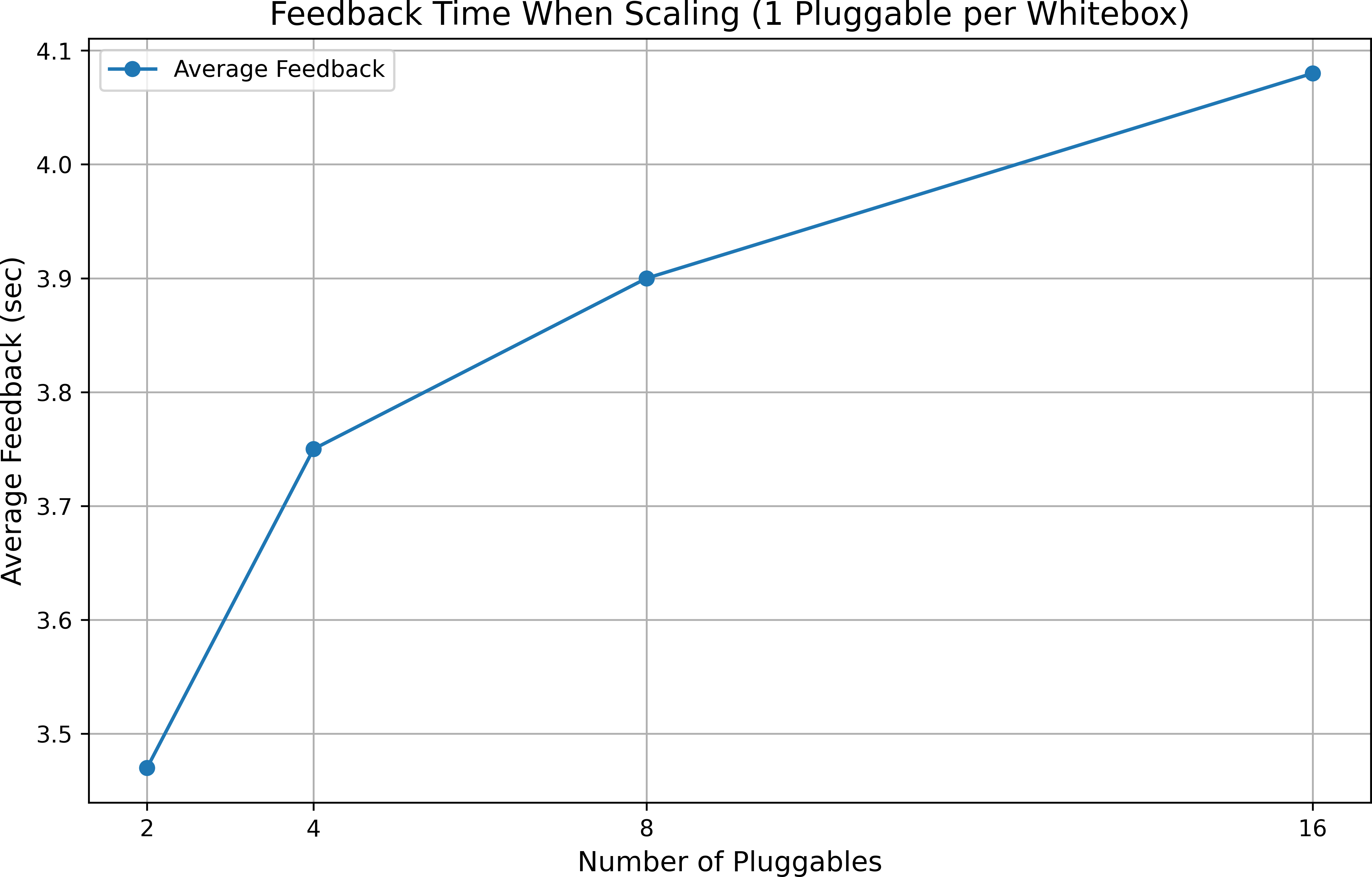}}
\caption{Scaling the topology.}
\label{figScaling}
\end{figure}

In Figure \ref{figScaling}, the scaling of the system is evaluated. Pluggables vary from 2 to 16 and the average feedback (laser configuration time) is depicted on the vertical axis. The average timestamp remains low even when the number of pluggables increases. In realtime, PacketCTL should be able to control multiple whiteboxes simultaneously for large transport networks. This way, the execution of efficient algorithms is facilitated in the ROADM part of the network as well.

\section{Conclusion}

The next generation of optical switching equipment offers a performance boost by transferring most of the computational overhead at the hardware layer. In the current setup, efficiency in the control plane increases in metro topologies due to the enabling of advanced configuration procedures for the pluggable devices, while utilising adaptive logic. This is occurring while the network operates under the supervision of a packet controller and multiple software agents that are executing at the whiteboxes. Also, the processing capabilities of dedicated chipsets boost performance since integral parts of a typical computer processing chain are bypassed. The decrease of the average laser configuration time is shown to be feasible even when the network topology is scaling.

\bibliographystyle{IEEEtran}
\bibliography{whitopen_conf}

\vspace{12pt}

\end{document}